\documentstyle[aps] {revtex}
\tighten
\twocolumn
\input{psfig.sty}
\begin{document}
\draft
\title{Absorption spectrum of 
a weakly n-doped semiconductor
quantum well} 
\author{F.X.~Bronold$^{1,2}$}
\address{$^1$Institut f\"ur Theoretische Physik, Otto-von-Guericke-Universit\"at
Magdeburg, D-39016 Magdeburg, Germany}
\address{$^2$Faculty of Engineering, Niigata University, Ikarashi 2-Nocho 8050, 
Niigata
950-2181, Japan}
\maketitle
\begin{abstract}
We calculate, as 
a function of temperature and conduction band
electron density, the optical absorption 
of a weakly n-doped, idealized semiconductor quantum well.
In particular, we focus on the absorption band 
due to the formation of a charged 
exciton. We conceptualize the
charged exciton as an itinerant excitation intimately linked
to the dynamical response of itinerant conduction band electrons 
to the appearance of the photo-generated valence band hole. 
Numerical results for the absorption in the vicinity of the
exciton line are presented and the 
spectral weights associated with,
respectively, 
the charged exciton band and the exciton line
are analyzed in detail. 
We find, in qualitative agreement with experimental data, 
that the spectral weight of the charged exciton  
grows with increasing conduction band electron density and/or 
decreasing temperature at the expense of the exciton.
\end{abstract}


Following Kheng et al.'s 
seminal experiment \cite{Kheng93}, which showed 
that in a weakly n-doped semiconductor quantum well (QW)
photo-generated excitons 
are accompanied by negatively charged excitons
\cite{Stebe89}, i.e. by bound states 
comprising two conduction band (CB) electrons and 
one valence band (VB) hole, numerous 
theoretical \cite{Stebe97,Stebe98,XminusTheo}
and experimental \cite{XminusExp,Eytan99,Manassen96} studies have been 
devoted to investigate optical properties of weakly n-doped  
QWs. There is also growing experimental 
interest in weakly p-doped
QWs where positively charged excitons have
been identified \cite{Xplus}. 
Among the most prominent open questions 
is the question about the microscopic structure
of the trion. (We focus here on n-doped QWs and use the term 
`trion' to 
refer to a 
negatively charged exciton.) In particular, it is 
still an unresolved issue, 
whether the trion is a localized, defect-like excitation
or an itinerant excitation, freely propagating 
in the QW plane.
There is experimental 
evidence that in {\it modulation doped} QWs
the trion is indeed trapped by the potential  
fluctutations of the remote donors \cite{Eytan99}. 
On the other hand, in {\it optically doped}, high-quality 
QWs, where trions 
have been also unambiguously observed (see, for 
example, Ref.
\cite{Manassen96}), remote donors are absent.
Moreover, potential fluctuations caused by
interface roughness, which could also act as a source for
localization, are presumably very small. 
Therefore, the trion in these samples 
is most likely freely propagating. Evidently, 
the broad theoretical task is to develop a trion
theory for {\it realistic} QWs treating both
possibilities on an equal footing, which is however rather 
formidable. As a first step, we focus therefore 
on an {\it idealized} QW, neglecting
localization processes altogether. 
In particular, we demonstrate, that in such an idealized, 
high-quality sample, it is the
{\it dynamical response} of itinerant CB electrons to the
appearance of the photo-excited VB hole, which, at low densities, 
inevitably gives rise to an itinerant trion and to the 
experimentally observed absorption band below the exciton line. 
In contrast to previous {\it few-body} theoretical studies
of the trion groundstate properties
\cite{Stebe89,Stebe97,XminusTheo}
and the trion absorption band \cite{Stebe98}, our genuine
{\it many-body} theoretical approach accounts for both
trion and exciton. Moreover,           
because the dynamical response of CB electrons is also of utmost 
importance  
at high density 
\cite{SchmittRink89}, our approach indicates how a   
complete lineshape theory for n-doped QWs, covering the 
whole density range, could be 
constructed. 

Since the details of the electronic band structure of a finite width 
QW unnecessarily masks the many-body theoretical
concepts we are envoking, we introduce our approach for  
a strictly two-dimensional (2D) idealized QW with two 
(isotropic) parabolic bands. 
Of course, with an appropriate  
choice of single particle states our formalism can be also 
applied to a more realistic QW model. 
Our calculation is based 
on linear response
theory which relates the (interband) optical 
susceptibility $\chi(\omega)$ to the $eh$-pair 
\cite{notation} propagator
$P(12;\omega)$ \cite{propagator}. 
Using techniques developed in nuclear many-body physics \cite{Schuck71},
$P(12;\omega)$ satisfies a Dyson equation  
(here `1' etc. stands for 2D momentum and
spin variables, see below however, and summation convention is implied),
\begin{eqnarray}
P(12;\omega)&=&
P^{(0)}(12;\omega)
+P^{(0)}(13;\omega)
\nonumber\\
&\times&[M^{st}(34)+\delta M(34;\omega)]
P(42;\omega),
\label{Dyson}
\end{eqnarray}
with an $eh$-pair self-energy containing a static part, $M^{st}(34)$,
and a dynamic part $\delta M(34;\omega)$. 
For a momentum independent transition matrix
element $r_{vc}$,
the optical susceptibility  
$\chi(\omega)=|r_{vc}|^2\sum_{12}
P(12;\omega)$,
the imaginary part of which determines the optical absorption,
i.e. apart from a constant factor 
$I_{abs}(\omega) \sim -Im\chi(\omega)$. 

\begin{figure}[t]
\hspace{2.0cm}
\psfig{figure=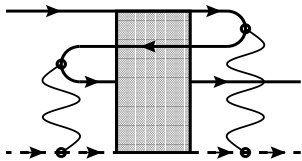,height=2.0cm,width=4.0cm,angle=0}
\vspace{0.5cm}
\caption[fig1]
{Diagrammatic representation of one term (out of four)
contributing to $\delta M(34;\omega)$. Solid lines in forward  
(backward) direction stand for $e$ ($\bar{e}$) propagation,
wavy lines depict Coulomb interactions and 
dashed lines denote $h$ propagation. In the low density limit 
the $\bar{e}$ line is free because scattering events involving a
conduction band hole are strongly suppressed by phase space
factors. The hatched box depicts the $eeh$ propagator $G$
to be determined in the Faddeev-Watson approximation \cite{Schuck71}.  
The remaining three diagrams have the same structure except that the 
first (last) scattering events are $eh$ ($ee$), $ee$ ($eh$), and
$ee$ ($ee$), respectively. Since we focus on the groundstate of the
$eeh$ cluster, we 
restrict the (intermediate) spin summations in these diagrams to 
anti-parallel electron spins. All spin summations can then be done
explicitly, yielding an overall factor of $2$ in the final expression
for $\chi(\omega)$.  
}
\label{fig1}
\end{figure}

The Dyson equation is valid for arbitrary CB electron densities. 
We are concerned, however, with low densities, where, 
in leading order, CB electrons do neither modify the 
wavefunction nor the energy of bound states (i.e. exciton and 
trion). 
Accordingly, we can ignore Pauli blocking as well as screening 
and treat CB electrons solely 
as a reservoir for bound state formation.    
We can, furthermore,
neglect (first order) electron and hole self-energy
corrections. Then,
\begin{eqnarray}
P^{(0)}(12;\omega)&=&
{\delta_{12}\over{\omega+i\eta-\epsilon_c(1)-\epsilon_v(-1)}},
\\
M^{st}(34)&=&-v(34),
\end{eqnarray}
with $\epsilon_{v/c}(1)$ and $v(34)$ denoting, respectively,
the bare 2D conduction and valence band
dispersion relations and the unscreened 2D Coulomb interaction.
The dynamical response of CB electrons, i.e. the creation of 
{\it virtual} 
$e\bar{e}$ pairs,
is contained in the dynamical part
$\delta M(34;\omega)$ which is related to a 
eight-point ($ee\bar{e}h$) function $R$.  
In the dilute limit 
we can show, 
however, using again techniques developed in Ref. \cite{Schuck71}, 
together with a 
low-density (or $\bar{e}$-line) expansion for $\delta M(34;\omega)$, 
that the eight-point
function R featuring in $\delta M(34;\omega)$ {\it factorizes}, 
in leading order of the CB 
electron density, into 
a two-point function
describing the (free) propagation of a CB hole and 
a six-point function $G$ describing the (correlated) propagation of an 
$eeh$ cluster (see Fig. \ref{fig1}). Hence, at low
densities, due to the creation of a ``single'' virtual 
$e\bar{e}$ pair,
optically generated $eh$ pairs are 
intrinsically coupled to 
$eeh$ clusters. 
If we then employ the spectral representation for $G$, 
keeping only bound $eeh$ clusters, which we expect to influence the 
optical response most dramatically, and anticipating that the
trion, i.e. the groundstate of the bound $eeh$ cluster, 
has anti-parallel electron spins,
we find  
(now a bold `{\bf 1}' etc. denotes only 2D momentum variables)
\begin{eqnarray}
\delta M({\bf 34};\omega)=
\sum_{{\bf 5}n} f_e({\bf 5})
{{\Xi_{{\bf 5}n}({\bf 3})\Xi_{{\bf 5}n}^*({\bf 4})}
\over{\omega+i\eta+\epsilon_c({\bf 5})-
\Omega_n({\bf 5})}},
\label{Self}
\end{eqnarray}
where all multiple $eeh$ scattering processes are encoded in
a vertex function
$\Xi_{{\bf 5}n}({\bf 3})=
\sum_{\bf 7} v({\bf 57})
\large[\Psi_{{\bf 5}n}({\bf 37})-\Psi_{{\bf 5}n}({\bf 7},{\bf 3+5-7})\large]$
given in terms of a  momentum-space $eeh$ wavefunction
$\Psi_{{\bf 5}n}({\bf 37})$, whose
Fourier transform to real space satisfies a
2D $eeh$ Schr\"odinger equation
with eigenvalue $\Omega_n({\bf 5})$.
Here, `n' and `{\bf 5}' depict, respectively, internal quantum numbers
and the (center of mass) momentum of
the (propagating) bound $eeh$ cluster. The 
Fermi function,
$f_e({\bf 5})$,
coming form the free propagation of the CB hole, 
accounts for the reservoir of CB electrons. 
Note, for vanishing CB electron density
$\delta M({\bf 34};\omega) \rightarrow 0$ and
Eqs. (\ref{Dyson}) -- (\ref{Self}) reduce to the 
ladder approximation for the exciton. 

To demonstrate that 
$\delta M({\bf 34};\omega)$ gives rise to the 
absorption band below the exciton 
line, we 
approximately solve
Eqs. (\ref{Dyson}) -- (\ref{Self}) for energies
$\omega$ close to the exciton groundstate energy $E_X$.
We adopt the following strategy: 
i) We keep in Eq. (\ref{Self}) only the groundstate of the
bound $eeh$ cluster,
i.e. the trion denoted by $n\equiv T$, whose
wavefunction and energy we determine with a variational technique
developed by St\'eb\'e and coworkers \cite{Stebe89,Stebe97}.
ii) We project
Eq. (\ref{Dyson}) to the exciton groundstate. 
Ignoring the coupling between the 
exciton groundstate and excited exciton states,
the projected $eh$-propagtor becomes 
$P_X(\omega)=\left[\omega+i\eta-E_X-\delta M_X(\omega) \right]^{-1}$, with 
an exciton self-energy  
$\delta M_X(\omega)=
\sum_{\bf 34} \phi_X^*({\bf 3})\delta M({\bf 34};\omega) \phi_X({\bf 4})$
(where $\phi_X({\bf 3})$ is the momentum-space wavefunction for the
exciton groundstate).
Evidently, this is only meaningful for large enough exciton binding
energies and, of course, only valid for $\omega$ in the 
vicinity of $E_X$. 
iii) Using the variational trion wavefuntion, we work out an analytical
expression for $\delta M_X(\omega)$ for non-degenerate CB electrons,
i.e. we assume low enough CB electron densities such that the 
Fermi energy is sufficiently 
inside the energy gap and that the Fermi function in Eq. 
(\ref{Self}) can be replaced by a Boltzmann function 
parametrized by the CB electron density \cite{HaugKoch90}.
For simplicity we also neglect the
weak dependence of
$\Xi_{\bf{5}n}({\bf 3})$ on
the trion momentum ${\bf 5}$.  

In terms of $P_X(\omega)$, the optical absorption 
in the vicinity of the exciton resonance reads
$I_{abs}(\omega) \sim 2 |r_{vc}|^2 |s_X|^2 Im P_X(\omega)$, 
with $s_X=\sum_{\bf 1}\phi_X({\bf 1})$. We now introduce
dimensionless quantities, measuring energy from the    
energy gap $E_g$ in units of the
2D exciton binding energy $4R$ ($R$ being the 
3D exciton Rydberg), and 
length in units of the 3D Bohr radius $a_B$. 
Specifically, we define $\tilde{\omega}=[\omega-E_g]/4R$ 
and combine 
all numerical factors into 
a constant $I_0$. Then,  
$I_{abs}(\tilde{\omega})=I_0 \cdot D_X(\tilde{\omega})$, 
with a dimensionless and, due to the spectral weight
sum rule \cite{NegeleOrland88}, 
normalized spectral (or lineshape) function
\begin{eqnarray}
D_X(\tilde{\omega})={1\over \pi} \cdot
{{\tilde{\eta}+\tilde{\Gamma}_X(\tilde{\omega})}\over
{[\tilde{\omega}+1-\tilde{\Delta}_X(\tilde{\omega})]^2+
[\tilde{\eta}+\tilde{\Gamma}_X(\tilde{\omega})]^2}}.
\label{lineshape}
\end{eqnarray}
Here $\tilde{\eta}=\eta/4R$ and the dimensionless functions 
$\tilde{\Delta}_X(\tilde{\omega})$ and $\tilde{\Gamma}_X(\tilde{\omega})$
are related to 
$\delta \tilde{M}_X(\tilde{\omega})\equiv\delta M_X(4R\tilde{\omega}+E_g)/4R$. 
Specifically,  
\begin{eqnarray}
\delta \tilde{M}_X(\tilde{\omega})
&=&\tilde{\beta}{M_T\over M_X}(na_B^2)
|u_T^X|^2 I(\tilde{\omega})
\\
&\equiv&\tilde{\Delta}_X(\tilde{\omega})-i\tilde{\Gamma}_X(\tilde{\omega}),
\label{Xself}
\end{eqnarray}
where $\tilde{\beta}=4\beta R$, $M_X$, $M_T$, and $n$ are, 
respectively, the thermal energy measured in 
units of $4R$, the exciton mass, the trion mass, and the
CB electron density. 
$u_T^X$ is the dimensionless exciton-trion coupling constant,
given 
by a real space
integral involving the variational trion
wavefunction, the Coulomb interaction, and the exciton 
wavefunction, and
\begin{eqnarray}
I(\tilde{\omega})=
\int_0^\infty dy 
{{e^{-\tilde{\beta}{M_T\over M_X}y}}\over{\tilde{\omega}+
i\tilde{\eta}-\tilde{\epsilon}_T+y}},
\label{ExpInt}
\end{eqnarray} 
with $\tilde{\epsilon}_T=[\Omega_T(0)-2E_g]/4R$ denoting the energy 
of a trion at rest, i.e. the
{\it internal} trion energy, which is the binding energy of {\it two} 
CB electrons to a VB hole. 
The assumption 
of non-degenerate 
CB electrons constrains the temperature and density
range in which our theory is valid. 
Specifically, using the well-known formula, 
$\beta\mu=\log[\exp(\pi n\beta/m_e^*)-1]$,
which relates the chemical potential $\mu$ of a 2D gas of free 
electrons (with mass $m_e^*$) to its density $n$, we estimate 
$\tilde{\beta}(na_B^2) < 1$, i.e. depending on the density the 
temperature has to be sufficiently high.  

\begin{figure}[t]
\psfig{figure=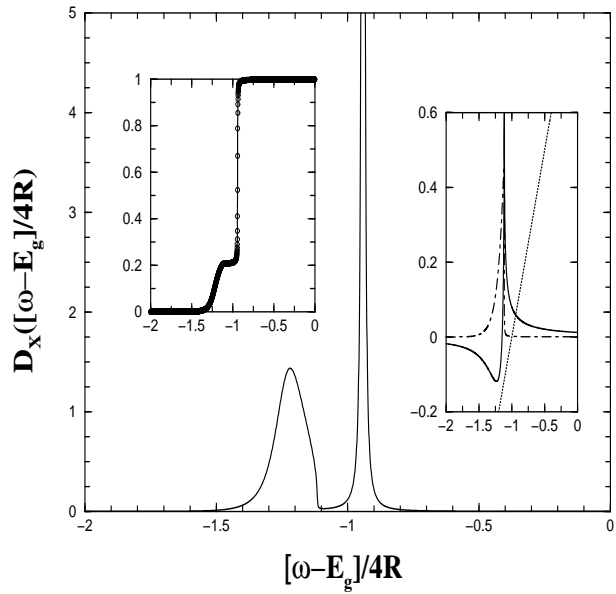,height=8.0cm,width=8.0cm,angle=-90}
\caption[fig2]
{The spectral function $D_X(\tilde{\omega})$ for 
$na_B^2=0.008$ and $4R\beta=10$. 
The effective mass ratio 
$\sigma=0.146$, corresponding to GaAs, and 
$\tilde{\eta}=0.001$. 
The left inset depicts the integrated spectral weight 
$C_X(\tilde{\omega})$,  
whereas the right inset 
displays $\tilde{\Gamma}_X(\tilde{\omega})$ 
(dot-dashed line) and
$\tilde{\Delta}_X(\tilde{\omega})$ (solid line). The energy   
for which $\tilde{\Delta}_X(\tilde{\omega})$ crosses 
$\tilde{\omega}+1$ (dotted line) defines the 
quasi-exciton line at $\tilde{\omega}=-0.94$. 
}
\label{fig2}
\end{figure}

We now discuss representative
numerical results for an effective mass
ratio $\sigma=m_e^*/m_h^*=0.146$ corresponding to GaAs. 
To obtain the exciton-trion
coupling constant and the internal trion energy,  
we specifically used, following St\'eb\'e and coworker 
\cite{Stebe89,Stebe97},
a 2D (22-term) Hylleraas variational wavefunction. 
Mathematical details will be presented elsewhere \cite{Bronold99}.
We obtained $u_T^X \sim 1.0896$ and $\tilde{\epsilon}_T=-1.1151$. 
Accordingly,
the binding energy (in units of $4R$)
for {\it one} electron  
$\tilde{W}_T=0.1151$ 
in agreement with previously obtained results \cite{Stebe89,XminusTheo}. 
The spectral function corresponding to 
$na_B^2=0.008$ and $\tilde{\beta}=10$
is shown in Fig. \ref{fig2}. 
The overall 
structure is in qualitative agreement with experimental results. 
Below a narrow peak at
$\tilde{\omega}= -0.94$, 
we notice a broad absorption band
characterized by 
a sharp high energy edge at
$\tilde{\omega}=-1.1151$ and a low-energy tail. 
To facilitate a microscopic understanding of the 
absorption features, we plot in the right inset of Fig. \ref{fig2}
real and imaginary part of the exciton self-energy.
The narrow peak is obviously a consequence of the pole 
of the spectral function $D_X(\tilde{\omega})$ at  
$\tilde{\omega}=-0.94\equiv\tilde{\epsilon}_X$,
where $\tilde{\epsilon}_X$ is the energy which simultaneously 
statisfies 
$\tilde{\Delta}_X(\tilde{\epsilon}_X)=\tilde{\epsilon}_X+1$
and $\tilde{\Gamma}_X(\tilde{\epsilon}_X)=0$. A pole in 
a spectral function is usually interpreted as  
a quasi-particle \cite{NegeleOrland88}. Accordingly, we   
attribute the narrow peak in Fig. \ref{fig2} to a  
{\it quasi-exciton}.   
We also see, the peak has actually no 
width, because, due to the assumed 
non-degeneracy of CB electrons, 
$\tilde{\Gamma}_X(\tilde{\epsilon}_X)=0$. 
The origin of the broad absorption band, in contrast, 
is the itinerant trion.
The sharp high energy 
edge, for example, comes from the pole of 
$\tilde{\Delta}_X(\tilde{\omega})$
at the trion formation energy $\tilde{\epsilon}_T=-1.1151$, 
i.e. the high energy edge of the band corresponds to a trion with 
momentum zero. The low energy tail, on the other hand,
is due to
trion states with finite momenta.
Note, due to the large value of $\tilde{\Gamma}_X(\tilde{\omega})$
in the vicinity of the edge, trion states with small 
momenta are significantly damped. Accordingly,   
the maximum of the absorption due to the trion is inside the 
band, corresponding to the creation of a trion with finite
momentum. The maximum
coincides with the high energy edge only
for very low CB electron density,  
in agreement with the calculations of Ref. \cite{Stebe98}. 
The spectral weight of the band, however, is then also very 
small (see Figs. \ref{fig3} and \ref{fig4} below). 
To analyze the spectral weights associated with,
respectively, the 
trion absorption band and the quasi-exciton line,  
we show in the
left inset of Fig. \ref{fig2} the integrated spectral 
weight up to an energy $\tilde{\omega}$ defined in terms of the cumulant
$C_X(\tilde{\omega})=\int_{-\infty}^{\tilde{\omega}} d\tilde{E}
D_X(\tilde{E})$. As can be seen,  
the integrated spectral weight of the trion absorption band corresponds
to the plateau value just below $\tilde{\omega}=-1$, i.e. 
$f_T=C_X(-1)$. The inset also demonstrates that the total 
spectral weight adds up to one, as it should, due to the spectral weight
sum rule. Accordingly, 
the spectral weight of the quasi-exciton is given by
$f_X=1-f_T$. 

We now present a quantitative investigation of the integrated spectral 
weights as a function of CB electron density 
and temperature. To that end, we calculated for various trion absorption 
bands the
integrated spectral weights $f_T=C_X(-1)$ and deduced from 
the sum rule, $f_X=1-f_T$, the corresponding spectral 
weights 
for the quasi-exciton. 
Fig. \ref{fig3} depicts typcial trion absorption bands
used to obtain the data  
shown in Fig. \ref{fig4}.
As expected, for fixed temperature, $f_T/f_X$
grows with CB electron density because the number of 
CB electrons available 
for trion formation increases. 
The density 
dependence is not linear, however. 
Instead,
we observe an initial fast increase and then a pronounced 
slowing down indicating the subtle interplay between trion and 
quasi-exciton. At high CB electron densities, beyond the 
validity range of our approach, we anticipate an even more  
complicated behavior. $f_T/f_X$ should show non-trivial 
features especially for densities close to a critical density, 
where the trion is expected to break up into   
an exciton and a free CB electron.
For fixed CB electron
density, $f_T/f_X$ increases with 
decreasing temperature. The increase also tends to saturate.
We emphasize, the numerical results  
for $f_T/f_X$ 
are consistent   
with experimental data, despite the simplicity of the
QW model, which neglects, e.g., the finite width of
the QW. 
Unfortuntately, experimentally achieved  
CB electron densities  
are above the density range in which
our theory is directly applicable.     
A quantitative comparison of, for example, lineshapes 
and oscillator strengths,  
requires, besides a more realistic QW model, to take modifications
of the exciton and trion wavefunctions and energies 
due to CB electrons explicitly into account. 
Our approach is feasible as long as bound states are 
stable. We only have to decorate 
various expressions with Pauli blocking 
factors, $\bar{f}_e(1)=1-f_e(1)$, and use a screened Coulomb 
interaction. The numerical implementation is
rather involved, however, and is beyond the scope of this
paper \cite{Bronold99}.

To summarize, we investigated
the optical absorption due to exciton and trion formation 
in a weakly n-doped idealized QW, conceptualizing 
the trion as an itinerant excitation formed in the course of the
{\it dynamical response} of itinerant CB 
electrons to the photo-generated VB hole. 
In particular, we focused, 
in an exploratory calculation, on low 
CB electron densities, treating CB electrons 
as a reservoir for bound state formation. 
The overall structure of the calculated
absorption is 
consistent with experimental data  
indicating that, at least in high-quality samples, the
dynamical response of CB electrons might play a key role
in the formation of trions. 

The author acknowledges useful discussions with Profs.\ H.\ B\"ottger 
and P.\ Schuck. 

%%%%%%%%%%%%%%%%%%%%%%%% REFERENCES %%%%%%%%%%%%%%%%%%%%%%%%%%%%%%%

%%%%%%%%%%%%%%%%%%%%%%%%% FIGURE CAPTIONS %%%%%%%%%%%%%%%%%%%%%%%%%%

\begin{figure}[t]
\psfig{figure=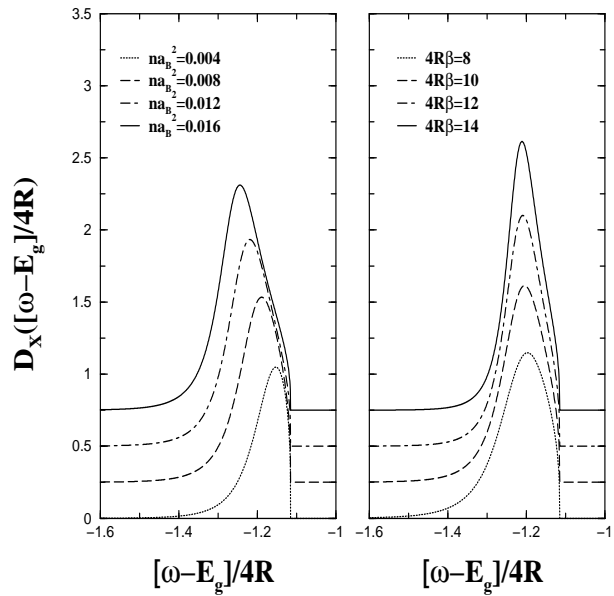,height=8.0cm,width=8.0cm,angle=-90}
\caption[fig3]
{The left and right panels show, respectively,
trion absorption bands
as a function of CB electron 
density for a fixed temperature $4R\beta=10$ and as a function
of temperature for a fixed  CB electron density $na_B^2=0.01$.
The quasi-exciton line located at $\tilde{\omega}>-1$
is not shown.
The effective mass ratio $\sigma=0.146$, corresponding
to GaAs, and
$\tilde{\eta}=0$. For clarity absorption bands corresponding to 
different
parameters are artificially shifted along the vertical axis.
}
\label{fig3}
\end{figure}

\begin{figure}[t]
\psfig{figure=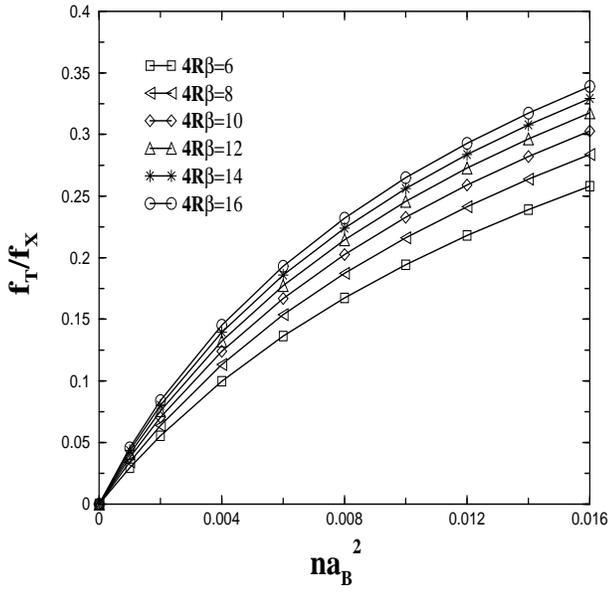,height=8.0cm,width=8.0cm,angle=-90}
\caption[fig4]
{The ratio of the integrated spectral weights ${f_T \over f_X}$
as a function of CB electron density and temperature.
The effective mass ratio $\sigma=0.146$, corresponding
to GaAs, and $\tilde{\eta}=0$. 
}
\label{fig4}
\end{figure}        


\begin{references}
\bibitem{notation}
$e$ and $\bar{e}$ denote, respectively, an electron and a hole in the
conduction band and $h$ stands for a hole in the valence band, i.e.
a trion is a bound $eeh$ complex.
\end{references}
\end{document}